\documentclass[aps,prd,onecolumn,groupedaddress,showpacs,nofootinbib,amssymb]{revtex4-2}
\usepackage[dvips]{graphicx}
\usepackage{amssymb}
\usepackage{amsmath}
\usepackage{graphicx,,color}
\usepackage{amsfonts}
\usepackage{bm}
\usepackage{cancel}
\usepackage{comment}
\usepackage{floatflt}

\newcommand\be{\begin{equation}}
\newcommand\ee{\end{equation}}

\allowdisplaybreaks[4]

\begin{document}

\title{Effects of the Axion Through the Higgs Portal on Primordial Gravitational Waves During the Electroweak Breaking}
\author{V.K. Oikonomou,$^{1,2}$}\email{voikonomou@gapps.auth.gr;v.k.oikonomou1979@gmail.com,voikonomou@auth.gr}
\affiliation{$^{1)}$Department of Physics, Aristotle University of
Thessaloniki, Thessaloniki 54124, Greece} \affiliation{$^{2)}$L.N.
Gumilyov Eurasian National University - Astana, 010008,
Kazakhstan}


 \tolerance=5000

\begin{abstract}
We investigate the effects of short axion kination eras on the
energy spectrum of the primordial gravitational waves
corresponding to modes that re-enter the Hubble horizon at the
post-electroweak symmetry breaking epoch, well within the
radiation domination era. Our main assumption is the existence of
an extremely weakly coupled hidden sector between the Higgs and
the axion, materialized by higher order non-renormalizable
dimension six and dimension eight operators, active at a scale $M$
of the order $20-100\,$TeV. This new physics scale $M$ which is
way higher than the electroweak scale, is motivated by the lack of
new particle observations in the large hadron collider to date,
beyond the electroweak scale. Once the electroweak symmetry
breaking occurs at $T\sim 100\,$GeV, the axion potential acquires
a new minimum due to the new terms generated by the electroweak
breaking, and the axion oscillations at the origin are
destabilized. In effect after some considerable amount of time,
the axion rolls swiftly to its new minimum, experiencing a short
kination epoch, where its energy density redshifts as $\rho_a\sim
a^{-6}$. After it reaches the new minimum, since the latter is
energetically less favorable that the Higgs minimum, it decays to
the Higgs minimum and the Universe is described again by the Higgs
minimum. The axion returns to the origin and commences again
oscillations initiated by quantum fluctuations, redshifts as dark
matter, and the same procedure is repeated perpetually. These
short axion kination eras may disturb the background total
equation of state parameter during the radiation domination era,
changing it from that of radiation $w=1/3$ to some value closer to
the kination value $w=1$. We examined the effect of a value
$w=1/2$ on the energy density of the primordial gravitational
waves. As we show, the energy spectrum of the gravitational waves
mainly depends on how many times the short axion kination epochs
occur, on the inflationary theory, the actual value of the
background equation of state parameter during the short kination
eras and finally on the reheating temperature. Our findings
indicate a characteristic shape in the energy spectrum which can
be observed in future gravitational wave experiments. We however
disregarded the contribution of the electroweak phase transition
on the gravitational waves, for simplicity and transparency of our
results.
\end{abstract}


\maketitle

\section{Introduction}

The focus in contemporary scientific research related to
theoretical and particle physics is undoubtedly in the sky and
specifically on gravitational wave experiments. The stage 4 Cosmic
Microwave Background (CMB) experiments
\cite{CMB-S4:2016ple,SimonsObservatory:2019qwx} and the future
gravitational wave experiments such as LISA, BBO, DECIGO and the
Einstein telescope
\cite{Hild:2010id,Baker:2019nia,Smith:2019wny,Crowder:2005nr,Smith:2016jqs,Seto:2001qf,Kawamura:2020pcg,Bull:2018lat,LISACosmologyWorkingGroup:2022jok}
will provide a solid answer on whether the inflationary era
\cite{inflation1,inflation2,inflation3,inflation4} took place, or
at least in the less optimistic scenario will further constrain
it. Although the stage 4 CMB experiments will directly probe the
$B$-mode seeds of inflation in the CMB temperature polarization
pattern, the future gravitational waves will probe the
inflationary modes, that is the tensor modes generated by the
curvature perturbations during inflation, that re-entered the
Hubble horizon shortly after the inflationary era, during the
reheating and the radiation domination era. These two eras are
hypothetical stages of our Universe's evolution, which are quite
mysterious themselves. During the radiation era it is theorized
that many hypothetical phenomena like the electroweak symmetry
breaking, have taken place. Regarding the latter, it provides an
elegant solution to the observed baryon asymmetry in the Universe,
since in many scenarios, the electroweak symmetry breaking stage
is followed by a strong first order phase transition, which is an
era of non thermal equilibrium, according to the Sakharov criteria
\cite{Sakharov:1967dj}. Apart from the baryon asymmetry, the
electroweak first order phase transition can in principle produce
primordial gravitational waves and these can be detected by future
interferometers
\cite{Apreda:2001us,Carrington:1991hz,Schabinger:2005ei,Kusenko:2006rh,McDonald:1993ex,Chala:2018ari,Davoudiasl:2004be,Baldes:2016rqn,Noble:2007kk,Zhou:2020ojf,
Weir:2017wfa,Hindmarsh:2020hop,Han:2020ekm,Child:2012qg,Fairbairn:2013uta,LISACosmologyWorkingGroup:2022jok,Caprini:2015zlo,Huber:2015znp,
Delaunay:2007wb,Chung:2012vg,Barenboim:2012nh,Curtin:2014jma,Child:2012qg,Senaha:2020mop,Grojean:2006bp,Katz:2014bha}.
These first order phase transitions are generated and further
assisted by singlet extensions of the Standard Model coupled only
to the Higgs particle
\cite{Profumo:2007wc,Damgaard:2013kva,Ashoorioon:2009nf,OConnell:2006rsp,Cline:2012hg,Gonderinger:2012rd,Profumo:2010kp,Gonderinger:2009jp,Barger:2008jx,
Cheung:2012nb,Alanne:2014bra,OConnell:2006rsp,Espinosa:2011ax,Espinosa:2007qk,Barger:2007im,Cline:2013gha,Burgess:2000yq,Kakizaki:2015wua,Cline:2012hg,
Enqvist:2014zqa,Barger:2007im}, or can be generated by Higgs self
couplings materialized by higher dimensional non-renormalizable
operators \cite{Chala:2018ari,Noble:2007kk,Katz:2014bha}.

Modified gravity in its various forms
\cite{reviews1,reviews2,reviews3,reviews4,reviews5} may
consistently describe the acceleration eras of our Universe, that
is the inflationary and the dark energy eras, and possibly all the
eras in between, without the need of scalar fields, however,
regarding the dark matter problem that still persists, one has to
be inceptive for its consistent description. Weakly interacting
massive particles (WIMP) still seem to be undetected, or well
hidden in a Standard Model hidden sector, thus currently the focus
and interest is on light mass particles, such as the axion
\cite{Preskill:1982cy,Abbott:1982af,Dine:1982ah,Marsh:2015xka,Sikivie:2006ni,Raffelt:2006cw,Linde:1991km,Co:2019jts,Co:2020dya,Barman:2021rdr,Marsh:2017yvc,Odintsov:2019mlf,Odintsov:2019evb,maxim,Anastassopoulos:2017ftl,Sikivie:2014lha,Sikivie:2010bq,Sikivie:2009qn,Masaki:2019ggg,Soda:2017sce,Soda:2017dsu,Aoki:2017ehb,Arvanitaki:2019rax,Arvanitaki:2016qwi,Machado:2019xuc,Tenkanen:2019xzn,Huang:2019rmc,Croon:2019iuh,Day:2019bbh,Oikonomou:2022ela,Oikonomou:2022tux,Odintsov:2020iui,Oikonomou:2020qah},
see also \cite{Semertzidis:2021rxs,Chadha-Day:2021szb} for reviews
and also an interesting simulation \cite{Buschmann:2021sdq} for
$\mu$eV range axions. Some experimental proposals are further
proposed recently for finding the axion \cite{BREAD:2021tpx}, and
also motivation for axions having mass of the order $m_a\sim
\mathcal{O}(10^{-10})\,$eV is provided by recent Gamma Ray Bursts
observations \cite{Hoof:2022xbe,Li:2022pqa}. In this article we
shall consider the effects of a direct non-renormalizable coupling
between the axion and the Higgs, on the energy density of the
primordial gravitational waves. Couplings between axions and Higgs
have been considered in the literature
\cite{Espinosa:2015eda,Im:2019iwd,Dev:2019njv} in a different
context, however in this paper we shall assume that the couplings
are in terms of higher order non-renormalizable operators. These
higher order non-renormalizable operators originate from a scale
$M$ way higher than the electroweak scale, a fact that is
motivated by the lack of new particle observations in the large
hadron collider (LHC)  beyond the electroweak breaking scale
\cite{Chala:2018ari}. When the electroweak symmetry breaking
occurs, nearly at a temperature $T\sim 100\,$GeV
\cite{Curtin:2014jma}, these higher order operators modify the
axion potential, and cause a new minimum to it. Eventually, the
axion oscillations at the origin of the axion potential are
destabilized after some time because these are unbounded, and the
axion is allowed to swiftly roll down to the new potential
minimum. Since the axion rolls swiftly in its new potential
minimum, it experiences a kination era and its energy density
scales as $\rho_a \sim a^{-6}$. With regard to the kination era
and its relation with light axion relics see for example
\cite{Ford:1986sy,Kamionkowski:1990ni,Grin:2007yg,Visinelli:2009kt}.
This short kination era may modify somewhat the background
equation of state (EoS) parameter from the radiation domination
value $w=1/3$ to a deformed value closer to the kination era value
$w=1$, depending on whether the axion composes all the dark matter
or some portion of it. Eventually when the axion reaches its new
minimum, the latter is energetically unfavorable compared to the
Higgs vacuum, and thus the axion minimum decays to the Higgs
minimum. Hence the actual vacuum of the Universe is the Higgs
minimum and the axion returns to the origin of its potential.
Quantum fluctuations generate the axion oscillations at the origin
since the dominant potential term for small field values is $\sim
\phi^2$, thus the axion starts to redshift again as dark matter.
After some considerable amount of time, the axion is again
destabilized due to the unbounded motion and starts to swiftly
roll its potential minimum again, and a new instant kination era
occurs for it. Thus the background EoS is changed again slightly,
and the procedure described is continued perpetually. It is thus
possible that during the post-electroweak breaking radiation
domination era, the Universe might have one or more deformations
of its background EoS $w=1/3$ to some value closer to the kination
era value $w=1$. We shall assume in a conservative  way that the
background EoS value during the kinetic phases of the axion is
$w=0.5$, and we shall investigate the effects of these axion
kinetic deformations on the energy spectrum of the primordial
gravitational waves. As we shall see, the effects can be
measurable in some scenarios, depending on the value of the total
background EoS parameter, the total number of these short kination
eras and finally on which inflationary scenario materializes the
inflationary era. Also we briefly discuss the effects of the axion
movement towards its new minimum in the case that the axion
initially slow-rolls towards it, and also we investigate the
effects of high temperature on the scenarios we discussed above.
This study however is purely academic, since the axion never
thermalizes with its background, thus the high temperature effects
are not justified. We included however the high temperature
effects just out of curiosity and for academic interest. It should
be mentioned that studies on the effects of stiff eras on
primordial gravity waves were performed in Refs.
\cite{Giovannini:1999qj,Giovannini:1999bh,Giovannini:1998bp}.

\section{Axion and the Higgs Portal via Dimension Six Operators and Dimension Eight Operators}

The axion scalar is the prominent dark matter candidate currently.
Its mass is likely in the sub-eV region, perhaps quite small,
ranging from $10^{-6}-10^{-27}$eV, according to different studies
\cite{Marsh:2015xka}. In this paper we shall consider the axion
evolution in the post-inflationary era, well inside the reheating
era and specifically during the electroweak breaking regime, which
occurs when the temperature of the Universe drops to $T\sim
100\,$GeV, after the Universe reached its high reheating
temperature. Regarding the reheating temperatures, it is still
questionable whether it was too high, but for the sake of the
argument we shall assume that the reheating temperature was in the
range $10^4-10^{7}\,$GeV, although there are studies that predict
a much lower reheating temperature. Also we shall assume that the
axion mass is of the order $m_a\sim \mathcal{O}(10^{-10})\,$eV
motivated by studies predicting such a mass based on Gamma Ray
Burst observations \cite{Hoof:2022xbe,Li:2022pqa}.

\begin{figure}[h!]
\centering
\includegraphics[width=20pc]{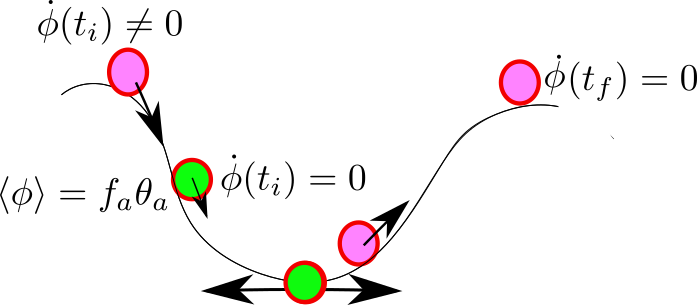}
\caption{The kinetic and ordinary misalignment axion models. In
the kinetic axion case, the axion has a large kinetic energy, and
the oscillations era is delayed.}
\label{pictorialrepresentationaxion}
\end{figure}

\begin{figure}
\centering
\includegraphics[width=20pc]{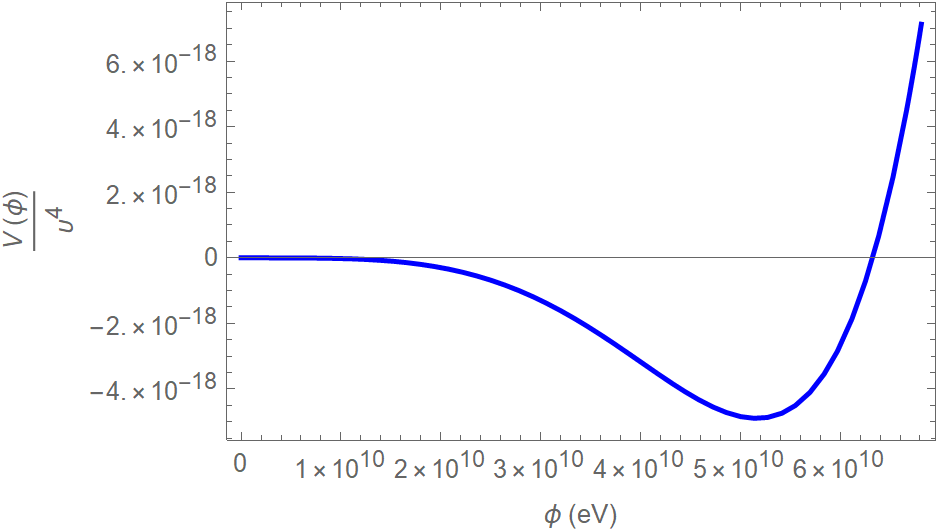}
\includegraphics[width=20pc]{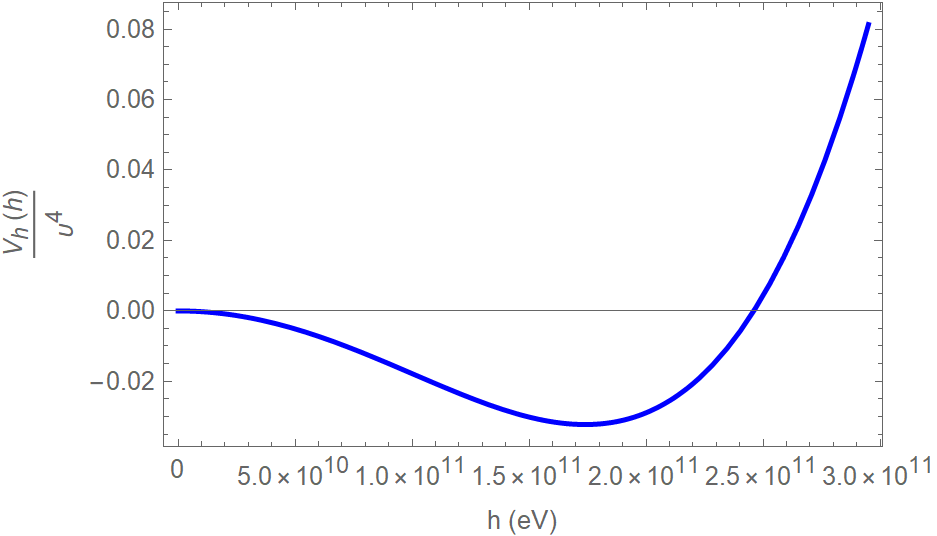}
\caption{The axion effective potential for $m_a\sim 10^{-10}$eV,
$M=20\,$TeV and $\lambda\sim \mathcal{O}(10^{-20})$ and $g\sim
\mathcal{O}(10^{-5})$ (left plot) and the Higgs potential after
electroweak symmetry breaking (right plot). The Higgs potential is
much deeper in its minimum value compared to the axion potential.
The axion vacuum is unstable compared to the Higgs
vacuum.}\label{higgsaxion68treeorder}
\end{figure}

We shall mainly be interested in the misalignment axion scenario
\cite{Marsh:2015xka,Co:2019jts}, in the context of which, the
primordial Peccei-Quinn $U(1)$ symmetry is broken during inflation
and the axion is misaligned from the vacuum state having a quite
large initial value $\phi_i\sim f_a$, where $f_a$ is the axion
decay constant, which will be assumed to be larger than
$f_a>10^{9}\,$GeV. The original axion potential has the form,
\begin{equation}\label{axionpotentnialfull}
V_a(\phi )=m_a^2f_a^2\left(1-\cos \left(\frac{\phi}{f_a}\right)
\right)\, .
\end{equation}
The term $\frac{\phi}{f_a}$ can quantify very well the
misalignment values of the axion, so when $\phi/f_a<1$, the axion
misalignment potential which describes the misalignment axion from
the minimum of the potential, is approximated as follows,
\begin{equation}\label{axionpotential}
V_a(\phi )\simeq \frac{1}{2}m_a^2\phi^2\, ,
\end{equation}
an approximation which recall it is valid when $\phi< f_a$ and is
expected no longer to hold true when $\phi\sim f_a$. This will
somewhat important at a later stage of our analysis. Regarding the
initial kinetic energy of the axion, there are two mainstream
scenarios which we shall take into account, the ordinary
misalignment axion scenario, according to which the axion starts
with zero kinetic energy and rolls slowly to the minimum of the
potential \cite{Marsh:2015xka}, and the second scenario is the
kinetic axion case \cite{Co:2019jts}. According to the latter
scenario, the axion starts with a large kinetic energy. In both
cases, when the Hubble rate of the Universe becomes of the order
of the axion mass $H\sim m_a$, the axion commences rapid
oscillations and redshifts as dark matter. However, this era of
oscillations is significantly delayed in the kinetic axion case,
because an era of kination follows after the quasi-de Sitter
inflationary era, and thus the axion has enough kinetic energy to
climb uphill its potential, and then starts to roll towards the
minimum of its potential. In the kinetic misalignment axion case,
the reheating temperature is presumably lower, compared to the
ordinary misalignment axion case, because the inflationary era is
prolonged by several $e$-foldings in the kinetic axion case
\cite{Oikonomou:2022tux,Oikonomou:2022ela}. The two scenarios are
described pictorially in Fig. \ref{pictorialrepresentationaxion}
We shall be interested in the effects of the electroweak symmetry
breaking on the axion evolution and the implications on the
primordial tensor modes crossing the horizon around that era. To
this end, we shall assume that some higher dimensional operators
of the axion are extremely weakly coupled to the Higgs sector
solely, while no other connection exists between the axion and the
Higgs sector. Thus we shall investigate the axion's evolution
effects on primordial gravitational waves through the Higgs portal
during the electroweak breaking epoch which is presumed to occur
at $T\sim 100\,$GeV \cite{Barenboim:2012nh,Curtin:2014jma}. From
the reheating temperature down to $100\,$GeV the Universe proceeds
without any changes with regard to its vacuum structure. The Higgs
portal will be constituted by six and eight dimensional operators
active at a scale $M$ which be assumed to be in the range $M\sim
20-100\,$TeV. Essentially, what we have is an extremely weakly
coupled theory originating at multiple TeVs scale. The proposed
Higgs-axion potential at tree order is the following,
\begin{equation}\label{axioneightsixpotential}
V(\phi,h)=V_a(\phi)-m_H^2|H|^2+\lambda_H|H|^4-\lambda\frac{|H|^2\phi^4}{M^2}+g\frac{|H|^2\phi^6}{M^4}\,
,
\end{equation}
with $V_a(\phi)$ being defined in Eq. (\ref{axionpotentnialfull}),
also the Higgs field prior to electroweak symmetry breaking is
$H=\frac{h+i h_1}{\sqrt{2}}$ being the Higgs scalar, also
$m_H=125$ GeV \cite{ATLAS:2012yve} is the Higgs boson mass,
$\lambda_H$ is the Higgs self-coupling, which are related as
$\frac{v}{\sqrt{2}}=\left(\frac{-m_H^2}{\lambda_H}\right)^{\frac{1}{2}}$,
where $v$ is the electroweak symmetry breaking scale
$v\simeq246\,$GeV. Furthermore, $m_a$ is the axion mass which will
be a free parameter in our theory, $M$ is the high scale of the
effective theory in which the non-renormalizable dimension six and
dimensions eight operator originate from. These dimension six and
dimension eight higher order non-renormalizable operators
originate from an effective theory active at the scale $M$, which
will be assumed to be way higher than the electroweak scale, of
the order $M=20-100\,$TeV, a fact that is further motivated by the
lack of new particle observations in the LHC, beyond the
electroweak breaking scale \cite{Chala:2018ari}. Also, $\lambda$
and $g$ are the dimensionless couplings or the Wilson coefficients
of the higher order effective theory of the axion to the Higgs,
which will be assumed to be small of the order $\lambda\sim
\mathcal{O}(10^{-20})$ and $g\sim \mathcal{O}(10^{-5})$. In this
paper we shall assume that the electroweak symmetry breaking
occurs and that a first order phase transition occurs in the
electroweak sector. This can be achieved in various ways, for
example via some scalar extension of the Higgs sector, only
coupled to the Higgs sector
\cite{Profumo:2007wc,Damgaard:2013kva,Ashoorioon:2009nf,OConnell:2006rsp,Cline:2012hg,Gonderinger:2012rd,Profumo:2010kp,Gonderinger:2009jp,Barger:2008jx,
Cheung:2012nb,Alanne:2014bra,OConnell:2006rsp,Espinosa:2011ax,Espinosa:2007qk,Barger:2007im,Cline:2013gha,Burgess:2000yq,Kakizaki:2015wua,Cline:2012hg,
Enqvist:2014zqa,Barger:2007im}, or by higher dimensional operators
of Higgs self couplings
\cite{Chala:2018ari,Noble:2007kk,Katz:2014bha}. The couplings of
the Higgs and axion particle ensure that the axion cannot affect
the electroweak sector, even after the electroweak breaking.
Before the breaking, the effects are well hidden due to the
weakness of the interaction. However, as we will see, the
electroweak breaking can affect the axion sector, causing the
axion to evolve differently after it occurs. Indeed, after the
electroweak symmetry breaking occurs at $T\sim 100\,$GeV, the
Higgs field acquires a vacuum expectation value, thus
$H=v+\frac{h+i h_1}{\sqrt{2}}$, therefore the axion potential
becomes modified, since the higher dimensional operators induce
self-interaction terms to the axion potential. Hence, the
resulting modified axion effective potential $\mathcal{V}_a(\phi)$
after electroweak breaking is,
\begin{equation}\label{axioneffective68}
\mathcal{V}_a(\phi)=V_a(\phi)-\lambda\frac{v^2\phi^4}{M^2}+g\frac{v^2\phi^6}{M^6}\,
,
\end{equation}
with $V_a(\phi)$ being defined in Eq. (\ref{axionpotentnialfull}).
These electroweak symmetry breaking induced self-couplings to the
axion, cause vacuum instability of the axion sector. In order to
pictorially see this vacuum instability in the axion sector let us
use the numerical values $m_a\sim 10^{-10}$eV, $M=20\,$TeV and we
take Wilson coefficients of the higher dimensional operators to be
of the order $\lambda\sim \mathcal{O}(10^{-20})$ and $g\sim
\mathcal{O}(10^{-5})$. In Fig. \ref{higgsaxion68treeorder} we plot
the axion effective potential (left plot) and the Higgs effective
potential (right plot). As it can be seen the total effective
potential in the Higgs minimum direction $(h,\phi)=(v,0)$ is much
deeper compared to the axion minimum $(h,\phi)=(0,v_s)$.
It is apparent that the dimension six and dimension eight
operators render the axion sector unstable, since the axion is now
allowed to acquire a vacuum expectation value. This could be
catastrophic if the axion is the only component of dark matter, or
is some considerable portion of the dark matter in the Universe.
But this is not so as we will see in the next section. Although
the Universe can have two physical vacua which can dominate, the
axion vacuum $(h,\phi)=(0,v_s)$ and the Higgs vacuum
$(h,\phi)=(v,0)$, the axion vacuum is unstable, and immediately
decays to the much deeper Higgs vacuum. We will discuss the
physical picture of the axion instability and its implications for
primordial gravitational waves in the next two subsections.

We need to note that even by including the 1-loop correction in
the axion effective potential,
\begin{equation}\label{oneloopaxionzerotemperature}
V^{1-loop}(\phi)=\frac{m_{eff}^4(\phi)}{64\pi^2}\left( \ln
\left(\frac{m_{eff}^2(\phi)}{\mu^2}\right)-\frac{3}{2}\right) \, ,
\end{equation}
does not change the overall physical picture, where
$m_{eff}^2(\phi)$ is,
\begin{equation}\label{axioneffectivemass}
m_{eff}^2(\phi)=\frac{\partial^2 V(\phi,h)}{\partial
\phi^2}=m_a^2-\frac{6 \lambda v^2 \phi^2}{M^2}+\frac{15 g v^2
\phi^4}{M^4}\, ,
\end{equation}
and $\mu$ denotes the renormalization scale. We have checked this
1-loop contribution for various values of the Wilson coefficients
for $M\sim 20-100\,$TeV and for values of the renormalization
scale $\mu$. The only issue is that the 1-loop contribution may
become complex for some field $\phi$ values and for some values of
the Wilson coefficients, which also indicates vacuum instability.
However, if temperature corrections were included, this imaginary
part would cancel. But the axion never thermalizes with the
particle environment, so one does not have to worry about high
temperature corrections. In a later section though, just for
academic curiosity, we shall include high temperature corrections
to see the effects in the physical picture. The result is the same
although high temperature delays the physics we shall describe.

\subsection{Metastable Vacuum Decay for Both Scenarios and Physical Description of the Resulting Phenomenology}

As we have seen in the previous subsection, after the electroweak
symmetry breaking in the Higgs sector, the dimension six and
dimension eight operators render the axion effective potential
unstable at the origin and it may develop a vacuum expectation
value. Thus after electroweak breaking, the Universe has two
competing vacua, the one corresponding to the Higgs field
$(h,\phi)=(v,0)$ and the one corresponding to the axion field
$(h,\phi)=(0,v_s)$. Which vacuum will dominates and which physics
will be materialized, depends on which is energetically favorable.
Depending on the values of the total axion-Higgs potential in the
two distinct vacua, the one vacuum may be highly unstable and can
tunnel quantum mechanically to the other vacuum according to the
Coleman description. The condition which determines which vacuum
is energetically favorable compares the depth of the effective
potential for the two distinct vacua, so in our case we have,
\begin{equation}\label{vacuumdecaycondition}
V(\phi,h)\Big{|}_{(h,\phi)=(0,v_s)}\gg
V(\phi,h)\Big{|}_{(h,\phi)=(v,0)}\, ,
\end{equation}
which essentially translates to the simple statement that the
``deeper'' vacuum energetically dominates over the other one, so
basically the energetically unfavorable vacuum is unstable and
decays to the stable vacuum.
\begin{figure}
\centering
\includegraphics[width=18pc]{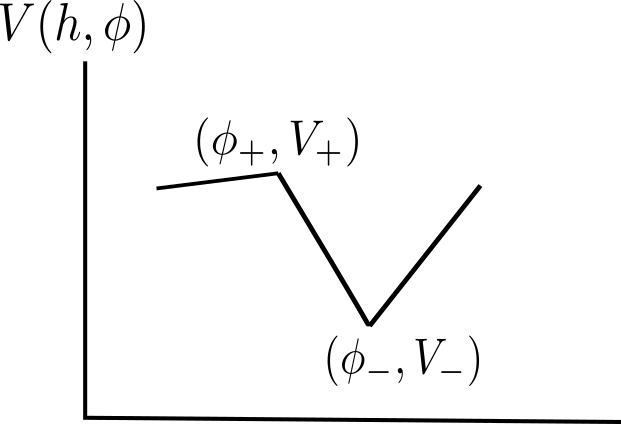}
\includegraphics[width=19pc]{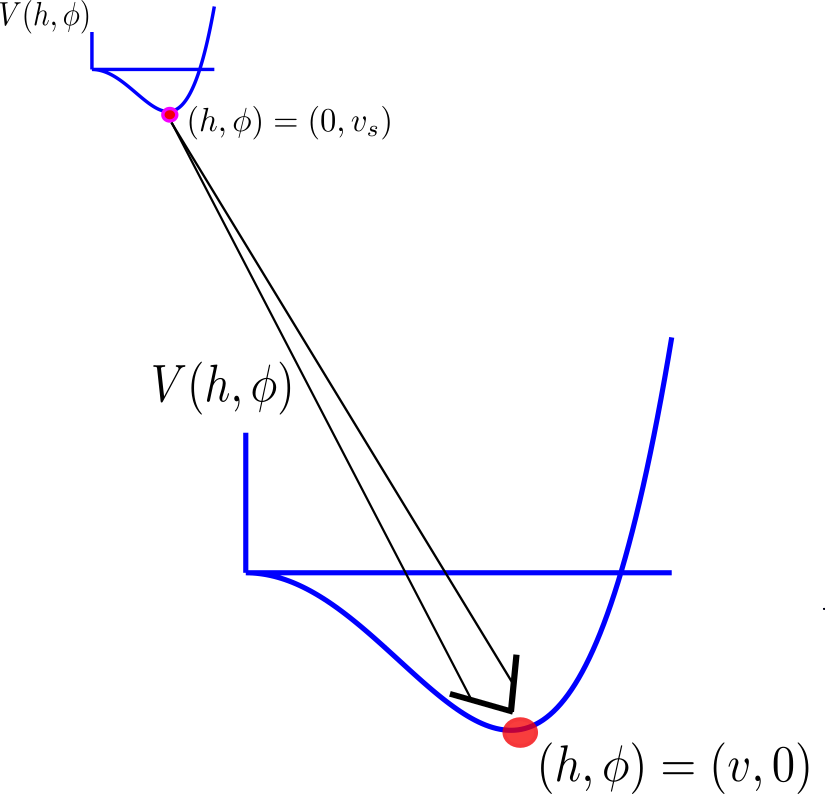}
\caption{Simplified triangular approximation for vacuum decay in
the axion-Higgs effective potential (left plot) and the vacuum
decay between the axion vacuum to the Higgs electroweak vacuum
(right plot).}\label{vacuumdecay}
\end{figure}
In our case, as it can be seen in Fig.
\ref{higgsaxion68treeorder}, the Higgs vacuum is by far more
energetically favorable compared to the axion vacuum, thus the
latter is highly unstable so it will decay eventually to the Higgs
vacuum. In order to quantify this picture, let consider the
triangular approximation used in Ref. \cite{Duncan:1992ai}, so
pictorially the physical situation is described in Fig.
\ref{vacuumdecay}, a simplified form of which is represented in
the left plot. In the triangular approximation, let $\phi_-$
denote the Higgs minimum $(h,\phi)=(v,0)$ of the axion Higgs
effective potential, $V_-$ the value of the effective potential at
the Higgs minimum $(h,\phi)=(v,0)$, $V_+$ and $\phi_+$ the
corresponding quantities for the axion minimum $(h,\phi)=(0,v_s)$.
Since in our case, the barrier connecting the two vacua in the
triangular approximation is practically elevated to the same level
as the axion minimum, the tunnelling rate parameter $B$ is,
\begin{equation}\label{decayrateparameter}
B=\frac{2\pi}{3}\frac{\Delta \phi^4 \tilde{\lambda}}{\Delta V}\, ,
\end{equation}
where in our case $\Delta \phi=\phi_+-\phi_-\simeq \phi_-$,
$\Delta V=V_+-V_-\simeq V_-$ and $\tilde{\lambda}$, which is the
steepness of the Higgs vacuum, is approximately of the order unity
$\tilde{\lambda}\sim \mathcal{O}(1)$. Thus the tunnelling rate
parameter $B$ which enters in the tunnelling rate $\Gamma\sim
e^{-B}$ of the axion vacuum $(h,\phi)=(0,v_s)$ to the Higgs vacuum
$(h,\phi)=(v,0)$, is a very small parameter of the order $B\sim
\mathcal{O}(10^{-42})$, thus the tunnelling rate is maximal and
the probability of tunnelling is of the order of unity. Hence the
axion vacuum is indeed highly metastable in our case and instantly
decays to the Higgs vacuum, in the way described in Fig.
\ref{vacuumdecay}. This is because the decay probability of a
metastable vacuum per unit time and volume $\gamma$ is
proportional to $\gamma\sim e^{-B}$. After a time average $T\sim
1/\gamma$, the metastable vacuum decays to the stable vacuum.
Since in our case, $\gamma$ takes the maximum value, the time
average is minimal and thus the axion vacuum instantly decays to
the Higgs vacuum.

\subsubsection{Metastable Vacuum Decay and the Evolution of the Axion Field}

We discussed the development of the axion minimum in the
axion-Higgs effective potential, we showed that the axion minimum
is unstable and it instantly decays to the Higgs minimum, so now
let us qualitatively discuss how the axion phenomenology is
affected by the metastable axion minimum and how does this effect
might affect the overall evolution of the Universe to some large
or small extent.

\begin{figure}
\centering
\includegraphics[width=19pc]{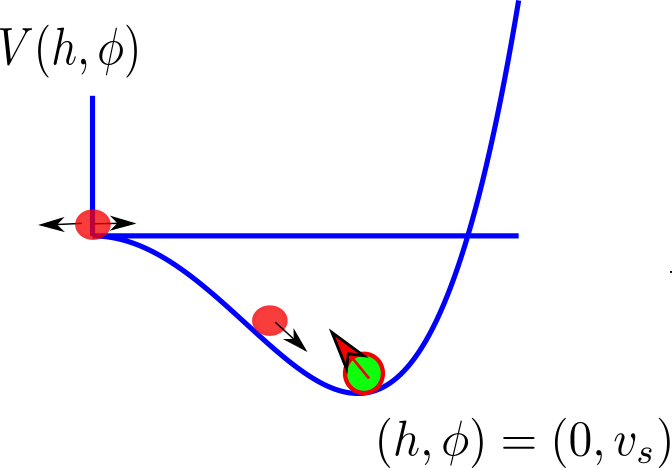}
\caption{The axion potential $V(\phi)$ after the electroweak
symmetry breaking develops a new minimum at the direction
$(h,\phi)=(0,v_s)$. The axion oscillations at the origin are
unbounded and the axion starts swiftly to roll down its potential,
to the new minimum. The axion undergoes a swift kination era for
which all the potential energy is transformed to kinetic energy.
Once it reaches the new minimum, the minimum decays to the
energetically favorable Higgs minimum $(h,\phi)=(v,0)$, and the
kinetic energy and the decay energy is transferred to axion which
returns to the origin. Quantum fluctuations generate again axion
oscillations at the origin, which after some considerable time
become unbounded and the same kination procedure is repeated
perpetually. }\label{pictorialkinationaxion}
\end{figure}

Firstly, exactly at the electroweak breaking epoch, the axion
potential develops a new minimum. Thus the axion, which prior to
electroweak breaking was performing oscillations around the
origin, now starts to feel the second minimum of the potential.
The oscillations of the axion at the origin are unstable, and thus
the origin becomes an unstable stationary point. Hence, eventually
after some time, which depends on the amplitude of the
oscillations and the initial conditions at the electroweak
breaking scale, the axion will start to roll down to its new
potential minimum, and its large potential energy-for the axion
standards-is converted to kinetic energy swiftly. It is thus
obvious that since the axion becomes effectively a kination fluid
for some small period of time, with that period of time being the
duration of the motion of the axion towards its new minimum. Hence
for this short period, the axion energy density scales as
$\rho_a\sim a^{-6}$, and by considering the fact that this occurs
after the electroweak transition, so basically deeply in the
radiation domination era, this effect might disturb significantly
the total background EoS of the Universe during the radiation
domination era, changing it from the value $w=1/3$ to some
different larger value, closer to the kination limit $w=1$, for
example $1/3<w<0.6$. When the axion reaches its new potential
minimum, the new axion vacuum decays instantly to the Higgs vacuum
and the axion returns to the origin of its effective potential.
Quantum fluctuations then will generate small oscillations about
the origin, since the axion potential for small $\phi$ values is
$\mathcal{V}_a(\phi)\sim \frac{1}{2}m_a^2\phi^2$, thus the axion
scales as dark matter $\rho\sim a^{-3}$ for the period that the
oscillations occur. When the oscillations amplitude grows larger,
the axion starts again to roll swiftly to the metastable axion
potential minimum, a new kination evolution for the axion occurs,
and the background EoS can be disrupted again during the radiation
domination era. Once the axion reaches swiftly the new minimum,
the metastable minimum decays and the axion again returns to the
origin. Thus this scenario may proceed until the radiation
domination era ends, and the Universe enters to the matter
domination era, and may continue even up to present day. This
behavior os pictorially described in Fig.
\ref{pictorialkinationaxion}
Therefore, in the case at hand, the axion energy density still
scales as cold dark matter $\rho_a\sim a^{-3}$, when the axion is
oscillating in the origin, but the evolution is swiftly
interrupted by very short kination eras. The latter last for a
very small amount of time, until the new axion minimum is reached.
Then the axion returns to the origin after the axion vacuum decays
to the Higgs vacuum. In the next section we shall consider the
effects of these short kination eras and how these may affect the
primordial gravitational waves. We need to note that only the
modes which re-enter the Hubble horizon during the radiation
domination era after the electroweak breaking will be affected.
This fact however strongly depends on how large the reheating
temperature was, thus the gravitational wave implications of these
short axion kination eras, which will disturb the radiation
domination era, should be considered in detail. This is considered
in the next subsection.

\subsection{Implications on Primordial Gravitational Waves}

In this section we shall discuss the general consequences of the
electroweak symmetry breaking on gravitational waves via the axion
sector. As we demonstrated in the previous section, the
electroweak symmetry breaking induces a new vacuum state for the
axion scalar field, to which the axion rolls rapidly once it gets
destabilized from the local maximum at the origin of its
potential.
\begin{figure}[h!]
\centering
\includegraphics[width=40pc]{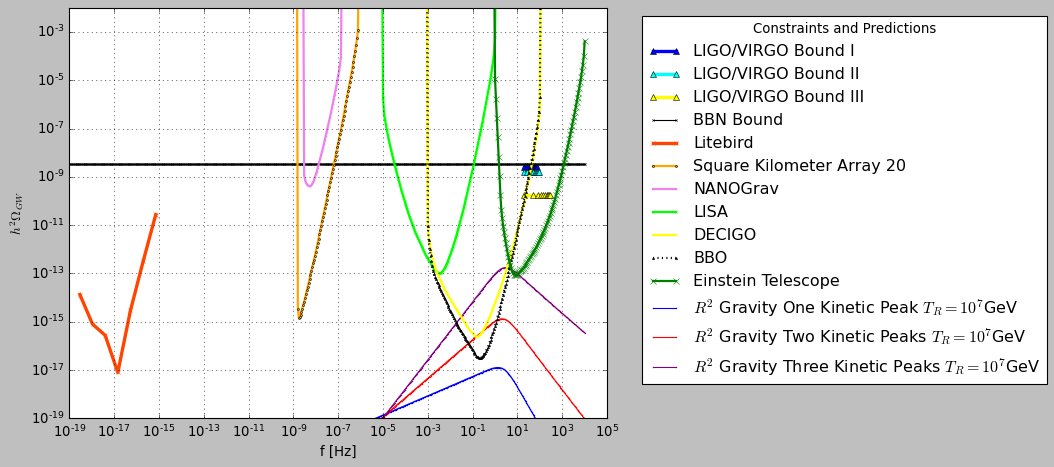}
\caption{The $h^2$-scaled gravitational wave energy spectrum for
the $R^2$-inflation canonical misalignment axion model, confronted
with the sensitivity curves of gravitational waves experiments for
the reheating temperature being high
$T_R=10^7\,$GeV.}\label{r2reheatingtemp107}
\end{figure}
This fast-roll may have measurable consequences for the background
total EoS of the Universe, since the axion solely experiences a
kination era and its energy density scales as $\rho_a\sim a^{-6}$,
thus it evolves slower than radiation. Thus if the axion composes
one part or all dark matter, this can affect the total EoS
parameter $w$, changing it from the standard radiation value
$w=1/3$ to some alternative higher value closer to the value $w=1$
which would indicate a kination era. If the axion is the sole dark
matter component, the total EoS parameter would be very close to
the kination value, but for demonstration purposes we shall assume
that the total EoS parameter changes to $w=0.5$. A larger value
than this proves to have more dramatic effects on the detection of
primordial gravitational waves, but we shall take a moderate path.
\begin{figure}[h!]
\centering
\includegraphics[width=40pc]{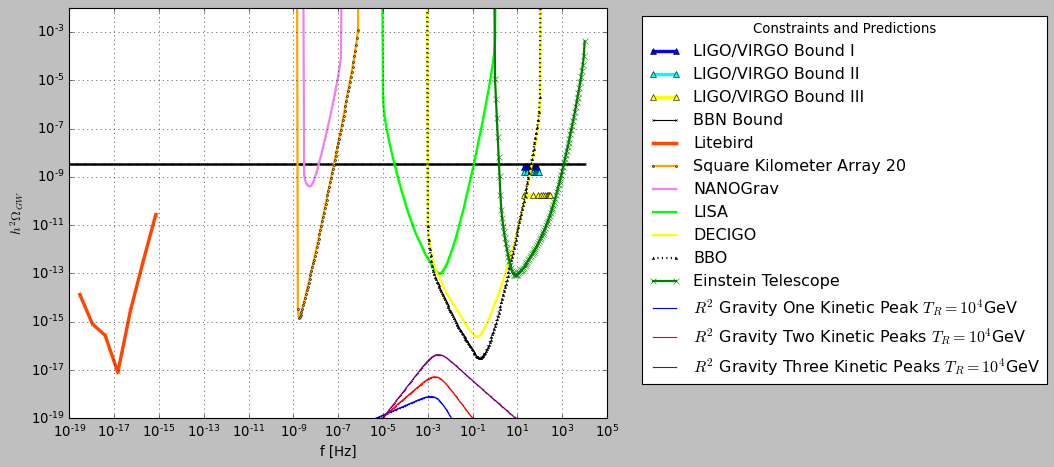}
\caption{The $h^2$-scaled gravitational wave energy spectrum for
the $R^2$-inflation canonical misalignment axion model, confronted
with the sensitivity curves of gravitational waves experiments for
the reheating temperature being low
$T_R=10^4\,$GeV.}\label{r2reheatingtemp104}
\end{figure}
Now regarding the axion rolling eras, these occur in a rapid way
and once the axion reaches the new minimum, the new axion vacuum
decays rapidly to the Higgs vacuum and thus the axion returns to
the origin of its potential to its initial vacuum state, and the
Universe is described by the vacuum $(h,\phi)=(v,0)$. However, due
to quantum fluctuations, the axion oscillations at the origin,
generated by the dominant part of the potential at the origin
$\sim \phi^2$, are destabilized again after some considerable
amount of time and the axion starts to roll again to the potential
minimum, since the other terms of the potential that cause the new
minimum are starting to dominate. The kination era occurs again
for the axion. This behavior continues perpetually for the axion,
with the latter experiencing instant kination eras during his
post-electroweak breaking evolution. This kind of behavior can
affect the last stages of the radiation domination era, relevant
for LISA, BBO and DECIGO physics, depending on how low the
reheating temperature was. In principle, no one can know how large
the reheating temperature was, there is motivation that it can be
as low as some hundreds MeV \cite{Hasegawa:2019jsa}. This in
principle could put in peril the electroweak phase transition,
however for the sake of the argument, we shall assume that the
reheating temperature can be either $T_R=10^7\,$GeV or
$T_R=10^4\,$GeV. The case of a lower reheating temperature could
be realized if for example a kinetic axion is controlling the
post-inflationary evolution, while larger reheating temperatures
can be realized by the canonical misalignment axion model (see
previous sections on this). The electroweak phase transition is
theorized to have occurred at approximately $T\sim 100\,$GeV, and
if assisted by scalar fields or by other mechanisms, it can be
first order. This could also generate another source of primordial
gravitational waves for this era which can work in a synergistic
way with the amplification of gravity waves due to the effects we
shall describe in this section, but for simplicity we omit the
primordial gravitational waves originating by the electroweak
phase transition. For a reheating temperature as high as
$T_R=10^7\,$GeV, the tensor modes re-entering the horizon at the
reheating temperature have a wavenumber of the order
$k_R=1.19\times 10^{15}$Mpc$^{-1}$. Hence at the electroweak phase
transition the relevant wavenumbers would be of the order $k_h\sim
10^{10}$Mpc$^{-1}$. Accordingly, for reheating temperature as high
as $T_R=10^4\,$GeV, the tensor modes re-entering the horizon at
the reheating temperature have a wavenumber of the order
$k_R=1.19\times 10^{12}$Mpc$^{-1}$. Hence, in this case at the
electroweak phase transition the relevant wavenumbers would be of
the order $k_h\sim 10^{9}$Mpc$^{-1}$ approximately. For academic
purposes and just from curiosity, we shall consider three
scenarios, first that during the radiation domination era there is
only axion kinetic epoch, secondly that there are two kination
epochs and thirdly three kination epochs. For the high reheating
epochs we shall assume that the kination epoch modes are for
example $k_{t_1}=6.5\times 10^{10}$Mpc$^{-1}$, $k_{t_1}=10\times
10^{10}$Mpc$^{-1}$ and $k_{t_1}=15\times 10^{10}$Mpc$^{-1}$ and
accordingly for the low reheating temperature, $k_{t_1}=6.5\times
10^{9}$Mpc$^{-1}$, $k_{t_1}=10\times 10^{9}$Mpc$^{-1}$ and
$k_{t_1}=15\times 10^{9}$Mpc$^{-1}$. Now, if the total EoS
parameter of the background is changed to $w$ at some relevant
wavenumber $k_{\mu}$, the $h^2$-scaled energy spectrum of the
primordial gravitational waves changes by a multiplication factor
$\sim \left(\frac{k}{k_{\mu}}\right)^{r_c}$, where
$r_c=-2\left(\frac{1-3 w}{1+3 w}\right)$
\cite{Gouttenoire:2021jhk}. Thus for three distinct such
deformations of the radiation domination era with $w=0.5$, one
should include three such multiplication factors to the
$h^2$-scaled energy spectrum of the primordial gravitational
waves. We shall denote this deformation of the kination eras as
$S_k(f)$, and the total $h^2$-scaled energy spectrum of the
primordial gravitational waves is
\cite{Kamionkowski:2015yta,Turner:1993vb,Boyle:2005se,Zhang:2005nw,Caprini:2018mtu,Clarke:2020bil,Smith:2005mm,Giovannini:2008tm,Liu:2015psa,Vagnozzi:2020gtf,Giovannini:2023itq,Giovannini:2022eue,Giovannini:2022vha,Giovannini:2020wrx,Giovannini:2019oii,Giovannini:2019ioo,Giovannini:2014vya,Giovannini:2009kg,Giovannini:2008tm,Kamionkowski:1993fg,Giare:2020vss,Zhao:2006mm,Lasky:2015lej,
Cai:2021uup,Odintsov:2021kup,Benetti:2021uea,Lin:2021vwc,Zhang:2021vak,Pritchard:2004qp,Khoze:2022nyt,Oikonomou:2022xoq,ElBourakadi:2022anr,Arapoglu:2022vbf,Giare:2022wxq}
(see also Ref. \cite{Odintsov:2022cbm} for a recent review),
\begin{align}
\label{GWspecfR}
    &\Omega_{\rm gw}(f)=S_k(f)\times \frac{k^2}{12H_0^2}r\mathcal{P}_{\zeta}(k_{ref})\left(\frac{k}{k_{ref}}
\right)^{n_T} \left ( \frac{\Omega_m}{\Omega_\Lambda} \right )^2
    \left ( \frac{g_*(T_{\rm in})}{g_{*0}} \right )
    \left ( \frac{g_{*s0}}{g_{*s}(T_{\rm in})} \right )^{4/3} \nonumber  \left (\overline{ \frac{3j_1(k\tau_0)}{k\tau_0} } \right )^2
    T_1^2\left ( x_{\rm eq} \right )
    T_2^2\left ( x_R \right )\, ,
\end{align}
where $k_{ref}=0.002$$\,$Mpc$^{-1}$ is the CMB pivot scale and
$S_k(f)$ is the contribution coming from the axion abrupt kination
eras. As it can be seen in the energy spectrum above, the
contributions from the inflationary era are also included. With
regard to the inflationary era, we shall consider two models that
can describe it, the $R^2$ model which yields a negative tensor
spectral index and an Einstein-Gauss-Bonnet theory which yields a
blue-tilted tensor spectrum. Specifically the $R^2$ model yields a
tensor spectral index $n_T=-0.000375$ and $r=0.003$, and the
Einstein-Gauss-Bonnet theory we shall consider is compatible with
the GW170817 even yielding a gravitational wave speed equal to
that of light's in vacuum, with the corresponding observational
indices being $n_T=0.37$ and $r=0.02$
\cite{Oikonomou:2022xoq,Oikonomou:2021kql,Odintsov:2020sqy}. Let
us briefly recall the theoretical frameworks of $R^2$ and
Einstein-Gauss-Bonnet gravity here, which yield the aforementioned
values for the inflationary observational indices. Regarding the
vacuum $F(R)$ gravity in the presence of a kinetic axion, which
does not affect at all the inflationary era, the action is,
\begin{equation}
\label{mainaction} \mathcal{S}=\int d^4x\sqrt{-g}\left[
\frac{1}{2\kappa^2}F(R)-\frac{1}{2}\partial^{\mu}\phi\partial_{\mu}\phi-V(\phi)+\mathcal{L}_m
\right]\, ,
\end{equation}
but recall that as it was shown in \cite{Oikonomou:2022tux} the
axion does not affect the inflationary era, just the duration of
it. Thus, the theory which mainly controls the dynamics of
inflation is the $F(R)$ gravity part,
\begin{equation}\label{effectivelagrangian2}
F(R)\simeq R+\frac{1}{M^2}R^2\, .
\end{equation}
Accordingly, the field equations become,
\begin{equation}\label{patsunappendix}
\ddot{H}-\frac{\dot{H}^2}{2H}+\frac{H\,M^2}{2}=-3H\dot{H}\, .
\end{equation}
and due to the slow-roll assumptions,
\begin{equation}\label{patsunappendix1}
-\frac{M^2}{6}=\dot{H}\, ,
\end{equation}
we get,
\begin{equation}\label{quasidesitter}
H(t)=H_I-\frac{M^2}{6} t\, ,
\end{equation}
The cosmological perturbations are quantified in a perturbative
way by the ``slow-roll'' indices
\cite{reviews1,Oikonomou:2022xoq},
\begin{equation}
\label{restofparametersfr}\epsilon_1=-\frac{\dot{H}}{H^2}, \quad
\epsilon_2=\frac{\ddot{\phi}}{H\dot{\phi}}\, ,\quad \epsilon_3=
\frac{\dot{F}_R}{2HF_R}\, ,\quad
\epsilon_4=\frac{\dot{E}}{2H\,E}\,
 ,
\end{equation}
with $E$ for the case at hand reads,
\begin{equation}\label{eparameter}
E=F_R+\frac{3\dot{F}_R^2}{2\kappa^2\dot{\phi}^2}\, .
\end{equation}
Following closely Ref. \cite{Oikonomou:2022tux} to which we refer
the reader for more details, by also taking into account the
kinetic axion at the end of inflation, the spectral index of
tensor and scalar perturbations at the end of inflation and the
tensor-to-scalar ratio take the form, $n_s\sim 1-\frac{2}{N}$,
$n_T\sim-\frac{1}{N^2}$ and $r\sim \frac{12}{N^2}$, where $N$ is
the $e$-foldings number. So for a kinetic axion theory $N$ must be
larger than 60 $e$-foldings, and we took 65 for our calculations.
Now regarding the Einstein-Gauss-Bonnet theory, the action is
\cite{Oikonomou:2022xoq,Oikonomou:2021kql,Odintsov:2020sqy},
\begin{equation}
\label{action} \centering
S=\int{d^4x\sqrt{-g}\left(\frac{R}{2\kappa^2}-\frac{1}{2}\partial_{\mu}\phi\partial^{\mu}\phi-V(\phi)-\frac{1}{2}\xi(\phi)\mathcal{G}\right)}\,
,
\end{equation}
where $R$ denotes the Ricci scalar, $\kappa=\frac{1}{M_p}$ where
$M_p$ is the reduced Planck mass. Also $\mathcal{G}$ stands for
the Gauss-Bonnet invariant which is
$\mathcal{G}=R^2-4R_{\alpha\beta}R^{\alpha\beta}+R_{\alpha\beta\gamma\delta}R^{\alpha\beta\gamma\delta}$
and $R_{\alpha\beta}$ and $R_{\alpha\beta\gamma\delta}$ denote the
the Ricci and Riemann tensor respectively.

\begin{figure}[h!]
\centering
\includegraphics[width=40pc]{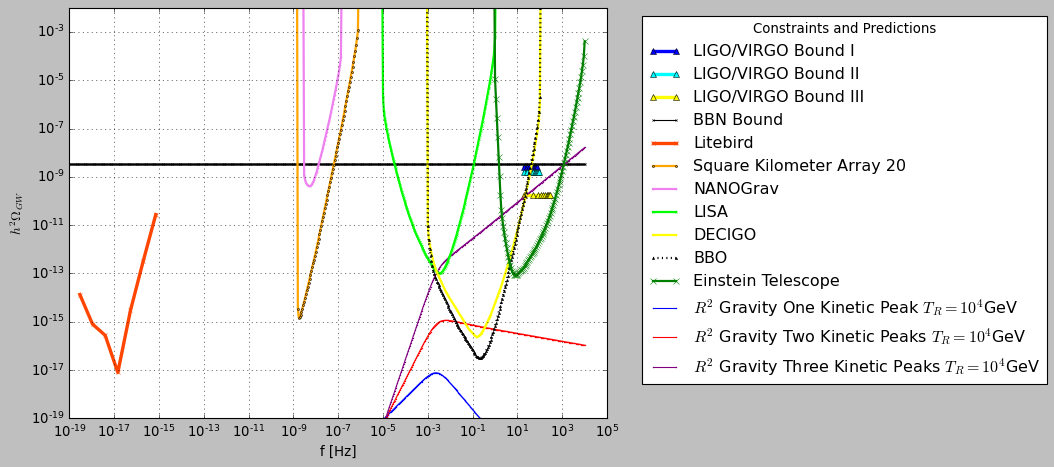}
\caption{The $h^2$-scaled gravitational wave energy spectrum for
the $R^2$ model, confronted with the sensitivity curves of
gravitational waves experiments for the reheating temperature
being low $T_R=10^4\,$GeV, with $w=0.9$.}\label{marginalr2}
\end{figure}

Following the argument of
\cite{Oikonomou:2022xoq,Oikonomou:2021kql,Odintsov:2020sqy}
regarding the gravitational wave speed being equal to unity, the
slow-roll indices of inflation are simplified, as follows
\cite{Oikonomou:2021kql},
\begin{equation}
\label{index1} \centering \epsilon_1\simeq\frac{\kappa^2
}{2}\left(\frac{\xi'}{\xi''}\right)^2\, ,
\end{equation}
\begin{equation}
\label{index2} \centering
\epsilon_2\simeq1-\epsilon_1-\frac{\xi'\xi'''}{\xi''^2}\, ,
\end{equation}
\begin{equation}
\label{index3} \centering \epsilon_3=0\, ,
\end{equation}
\begin{equation}
\label{index4} \centering
\epsilon_4\simeq\frac{\xi'}{2\xi''}\frac{\mathcal{E}'}{\mathcal{E}}\,
,
\end{equation}
\begin{equation}
\label{index5} \centering
\epsilon_5\simeq-\frac{\epsilon_1}{\lambda}\, ,
\end{equation}
\begin{equation}
\label{index6} \centering \epsilon_6\simeq
\epsilon_5(1-\epsilon_1)\, ,
\end{equation}
where $\mathcal{E}=\mathcal{E}(\phi)$ and $\lambda=\lambda(\phi)$
are defined as follows,
\begin{equation}\label{functionE}
\mathcal{E}(\phi)=\frac{1}{\kappa^2}\left(
1+72\frac{\epsilon_1^2}{\lambda^2} \right),\,\, \,
\lambda(\phi)=\frac{3}{4\xi''\kappa^2 V}\, .
\end{equation}
and hence the observational indices of inflation which are defined
as,
\begin{equation}
\label{spectralindex} \centering
n_{\mathcal{S}}=1-4\epsilon_1-2\epsilon_2-2\epsilon_4\, ,
\end{equation}
\begin{equation}\label{tensorspectralindex}
n_{\mathcal{T}}=-2\left( \epsilon_1+\epsilon_6 \right)\, ,
\end{equation}
\begin{equation}\label{tensortoscalar}
r=16\left|\left(\frac{\kappa^2Q_e}{4H}-\epsilon_1\right)\frac{2c_A^3}{2+\kappa^2Q_b}\right|\,
,
\end{equation}
where $c_A$ is the sound speed,
\begin{equation}
\label{sound} \centering c_A^2=1+\frac{Q_aQ_e}{3Q_a^2+
\dot\phi^2(\frac{2}{\kappa^2}+Q_b)}\, ,
\end{equation}
and with,
\begin{align}\label{qis}
& Q_a=-4 \dot\xi H^2,\,\,\,Q_b=-8 \dot\xi H,\,\,\,
Q_t=F+\frac{Q_b}{2},\\
\notag &  Q_c=0,\,\,\,Q_e=-16 \dot{\xi}\dot{H}\, ,
\end{align}
can be further simplified as follows,
\begin{equation}\label{tensortoscalarratiofinal}
r\simeq 16\epsilon_1\, ,
\end{equation}
\begin{equation}\label{tensorspectralindexfinal}
n_{\mathcal{T}}\simeq -2\epsilon_1\left ( 1-\frac{1}{\lambda
}+\frac{\epsilon_1}{\lambda}\right)\, ,
\end{equation}
Using the following model,
\begin{equation}
\label{modelA} \xi(\phi)=\beta  \exp \left(\left(\frac{\phi
}{M}\right)^2\right)\, ,
\end{equation}
where $\beta$ is a dimensionless parameter, and $M$ denotes a free
parameter with mass dimensions $[m]^1$. For this model, the
observational indices become,
\begin{align}\label{spectralpowerlawmodel}
& n_{\mathcal{S}}\simeq -1-\frac{\kappa ^2 M^4 \phi
^2}{\left(M^2+2 \phi ^2\right)^2}+\frac{4 \phi ^2 \left(3 M^2+2
\phi ^2\right)}{\left(M^2+2 \phi ^2\right)^2}\\ & \notag
+\frac{4608 \beta ^2 \phi ^6 e^{\frac{2 \phi ^2}{M^2}} \left(6
\gamma  \phi ^2+16 \beta  e^{\frac{\phi ^2}{M^2}} \left(M^2+\phi
^2\right)+9 \gamma  M^2\right)}{\left(M^2+2 \phi ^2\right)^4
\left(3 \gamma +4 \beta  e^{\frac{\phi ^2}{M^2}}\right)^3} \, ,
\end{align}
and
\begin{align}\label{tensorspectralindexpowerlawmodel}
& n_{\mathcal{T}}\simeq \frac{\phi ^2 \left(-4 \beta e^{\frac{\phi
^2}{M^2}} \left(M^4 \left(3 \kappa ^2 \phi ^2-2\right)+\kappa ^2
M^6-8 M^2 \phi ^2-8 \phi ^4\right)-3 \gamma \kappa ^2 M^4
\left(M^2+2 \phi ^2\right)\right)}{\left(M^2+2 \phi ^2\right)^3
\left(3 \gamma +4 \beta  e^{\frac{\phi ^2}{M^2}}\right)}
 \, .
\end{align}
while the tensor-to-scalar ratio reads,
\begin{equation}\label{tensortoscalarfinalmodelpowerlaw}
r\simeq \frac{8 \kappa ^2 M^4 \phi ^2}{\left(M^2+2 \phi
^2\right)^2}\, .
\end{equation}
For $\mu=[22.09147657871,22.09147657877]$, $\beta=-1.5$,
$\gamma=2$, and for $N=60$ $e$-foldings we obtain,
$n_{\mathcal{T}}=[0.378856,0.379088]$ and we will use this model
for the primordial gravitational waves predictions of this theory.
\begin{figure}[h!]
\centering
\includegraphics[width=40pc]{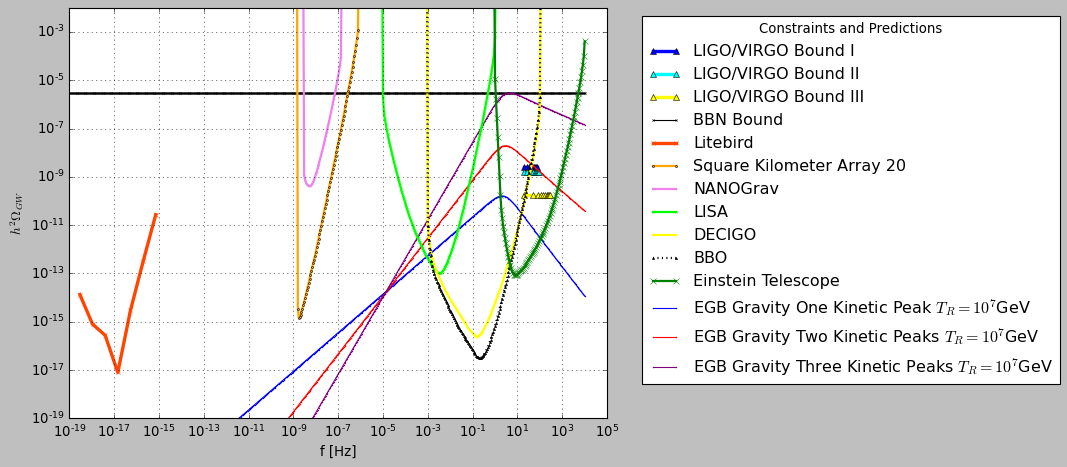}
\caption{The $h^2$-scaled gravitational wave energy spectrum for
the Einstein-Gauss-Bonnet model, confronted with the sensitivity
curves of gravitational waves experiments for the reheating
temperature being high
$T_R=10^7\,$GeV.}\label{egbreheatingtemp107}
\end{figure}
Now let us confront the corresponding theories with the
sensitivity curves coming from all future and present (NANOGrav)
gravitational wave experiments, taking also into account the
abrupt short kination eras of the axion sector, with one, two and
three kinetic eras occurring during the radiation domination era.
In our analysis we shall include the latest constraints coming
from the LIGO/VIRGO collaboration \cite{KAGRA:2021kbb} which
indicate that the energy spectrum of the primordial gravitational
waves must be $\Omega_{GW}\leq 5.8\times 10^{-9}$ for a flat and
frequency independent background for frequencies in the range
$(20-76.6)$Hz (LIGO VIRGO BOUND I), and furthermore
$\Omega_{GW}\leq 3.4\times 10^{-9}$ for a spectral index of $2/3$
power-law background in the range $(20-90.6)$Hz (LIGO VIRGO BOUND
II) and $\Omega_{GW}\leq 3.9\times 10^{-10}$ when the spectral
index is $3$, in the range $(20-291.6)$Hz (LIGO VIRGO BOUND III).
Moreover, the BBN bound is included \cite{Kuroyanagi:2014nba}.
\begin{figure}[h!]
\centering
\includegraphics[width=40pc]{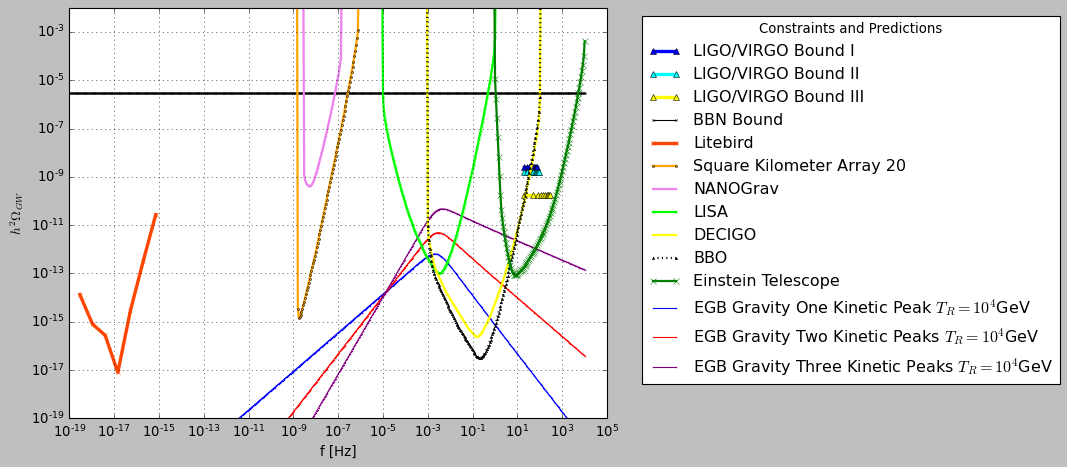}
\caption{The $h^2$-scaled gravitational wave energy spectrum for
the Einstein-Gauss-Bonnet model, confronted with the sensitivity
curves of gravitational waves experiments for the reheating
temperature being low $T_R=10^4\,$GeV.}\label{egbreheatingtemp104}
\end{figure}
In Figs. \ref{r2reheatingtemp107} and \ref{r2reheatingtemp104} we
present the $R^2$ model results with one two and three kination
peaks, while in Figs. \ref{egbreheatingtemp107} and
\ref{egbreheatingtemp104} we present the Einstein-Gauss-Bonnet
model results with again one, two and three kinetic peaks, for two
distinct reheating temperatures, namely $T_R=10^7\,$GeV and
$T_R=10^4\,$GeV. Now let us discuss the resulting phenomenology
and we start with the $R^2$ model. For a small reheating
temperature, the primordial gravitational wave energy spectrum
remains undetectable, however for a large reheating temperature
with three and two kination peaks can be detected by the BBO and
DECIGO experiments. In the Einstein-Gauss-Bonnet case, the
primordial gravitational waves are detectable even for a reheating
temperature as  low as $T_R=10^4\,$GeV and can be detected from
LISA, DECIGO and BBO, even for one kination peak. In fact for a
large reheating temperature, the two and three kination peaks are
already excluded by the LIGO/VIRGO constraints, as it can be seen
in Fig. \ref{egbreheatingtemp107}. We should note that we did not
include in our study the synergistic effect of the electroweak
phase transition, which if it is of first order, this could affect
the energy spectrum when the temperature of the Universe is
$100\,$GeV and could further amplify the spectrum. We solely
considered the effects of the axion kination eras, for simplicity,
in order to discriminate the effects of the kination eras. Before
closing, in order to stress the effect of the background EoS
parameter $w$ which we assumed that it is equal to $w=0.5$. If
this is closer to the kination value, for example $w=0.9$, then
the effects on the primordial gravitational wave energy spectrum
are amplified, see for example Fig. \ref{marginalr2}. However, it
is highly unlikely that the background EoS will be modified to
such extent during the radiation domination eras, by the short
kination eras of the axion.

\subsubsection{Brief Comment: Metastable Vacuum Decay During Matter Domination Era and Early Dark Energy}

Let us briefly discuss the implications of the rolling axion
towards to its new minimum during the matter domination era. In
principle this rolling era can have significant effects on the
matter domination era, changing the background EoS parameter from
the matter dominating value to something different. The most
interesting scenario, is the case that in which the axion does not
roll in a rapid way, but initially slow-rolls to its minimum.
Since in the matter domination era, the axion plays an important
role in the composition of dark matter, an initial slow-roll era
might change the background EoS to be de Sitter or even
quintessential. Thus it is possible that an early dark energy era
is realized in the context of our scenario, and basically not only
one, but multiple short dark energy eras are realized, depending
on the duration of the unstable axion oscillations at the origin.
These early dark energy eras play no role phenomenologically for
gravitational wave physics, but can explain the $H_0$-tension, see
for example \cite{Kamionkowski:2022pkx}. Of course the axion
rolling will change the background EoS parameter from $w=0$ to
$w=1$ in an abrupt way, so swift kination effects during the
matter domination era will take place too. We do not further
pursuit this aspect, since it is out of the purposes of this work.

\section{An Exotic Scenario: Tachyonic Instabilities Induced in the Axion Sector via the Higgs Portal}

Now we shall consider an exotic scenario according to which the
Higgs portal can induce tachyonic instabilities in the axion
sector. This is rather unappealing phenomenologically, but we
shall consider this for completeness. It should be noted that this
mechanism can in principle apply to any other Standard Model
singlet axion-like-particle coupled to the Higgs and that belongs
to a hidden sector. In this case, once the axion rendered a
tachyon, it can no longer act as dark matter. Thus if the axion is
the main component of cold dark matter, this is a destructive
scenario for the axion dark matter perspective Universe. This
scenario can only apply in the case that the axion is some part of
cold dark matter, but not all of it.

The axion-Higgs potential we shall consider in this case is,
\begin{equation}\label{higgsaxiontreepotential}
V(H,\phi)=-m_H^2|H|^2+\lambda_H|H|^4+\frac{1}{2}m_a^2\phi^2-\lambda|H|^2\phi^2+\frac{g}{M^2}\phi^4
|H|^2\, ,
\end{equation}
with, $H=\frac{h+i h_1}{\sqrt{2}}$ being the Higgs scalar, also
$m_H=125$ GeV is the Higgs boson mass, $\lambda_H$ is the Higgs
self-coupling, which are related as
$\frac{v}{\sqrt{2}}=\left(\frac{-m_H^2}{\lambda_H}\right)^{\frac{1}{2}}$,
where $v$ is the electroweak symmetry breaking scale $v\simeq
246\,$GeV. Furthermore, $m_a$ is the axion mass which will be a
free parameter in our theory, $M$ is the high scale of the
effective theory in which the non-renormalizable dimension six
operator originates from, and $\lambda$ and $g$ are the coupling
of the axion to the Higgs which will be assumed to be extremely
small of the order $\lambda\sim \mathcal{O}(10^{-43})$ and $g\sim
\mathcal{O}(10^{-35})$, with $g$ being the Wilson coefficient of
the higher dimension operator. Thus the renormalizable Higgs
portal axion interaction is extremely weak and also the effective
axion-Higgs theory at the scale $M$ is also extremely weakly
coupled. The weakly coupled nature of the proposed theory
generated at high TeV scales, is essentially a safe path for Higgs
phenomenology, since the branching fraction of the Higgs to the
hidden axion sector is severely constrained \cite{Chung:2012vg}.
Specifically if the branching fraction of the Higgs to invisible
axion particles $\mathrm{BR}_{inv}$ via $h\to \phi \phi$ would be
large, then it would be hard to detect the Higgs in the LHC
\cite{Chung:2012vg}. Also the upper bounds of the branching
fraction of the Higgs to the invisible hidden axion sector are
$\mathrm{BR}_{inv}<0.30-0.75$ at $95\%$CL \cite{Chung:2012vg}. The
potential of Eq. (\ref{higgsaxiontreepotential}) has an inherent
$Z_2$ symmetry $\phi\to -\phi$ which renders the axion particle
stable towards decays, since there is no mixing between the Higgs
and the axion, and also the axion is considered to be a Standard
Model singlet. Such $Z_2$ singlet-Higgs models are used frequently
in the literature  and can be related to dark matter components
\cite{Barger:2007im,Chung:2012vg}.

A mentionable feature of the renormalizable Higgs-axion
interaction $-\lambda|h|^2\phi^2$ is the negative sign we chose in
it. Such negative couplings are particularly favored  by Higgs
phenomenology. Indeed, the excess of the Higgs decay to diphotons
is particularly enhanced  for such a negative coupling
\cite{Chung:2012vg}. Needless to say that the $h\to \gamma \gamma$
channel is very sensitive to new physics, thus the choice for a
negative coupling is well-motivated. Considering the
renormalizable interaction $-\lambda|h|^2\phi^2$, the decay rate
of the Higgs to the hidden invisible axion sector is in our case
\cite{Barger:2007im},
\begin{equation}\label{axionhiggsdecayrate}
\Gamma (h\to \phi \phi)=\frac{\lambda^2v^2}{32\pi
m_H}\sqrt{1-\frac{4m_a^2}{m_H^2}},
\end{equation}
so with the axion mass being at the sub-eV range and with the
coupling of the renormalizable Higgs-axion interaction being of
the order $\lambda\sim \mathcal{O}(10^{-43})$, the decay rate of
the Higgs to the hidden axion sector is practically zero. Thus,
the branching fraction of the Higgs to Standard Model particles is
not affected at all in our case. A large decay rate would affect
and practically would reduce the strength of the Higgs boson
signal at the LHC, as we already mentioned, but this is not our
case.

A confusing question that is raised for the coupling $\lambda$
entering in the renormalizable Higgs-axion interaction
$-\lambda|h|^2\phi^2$ is whether $\lambda$ is allowed to take such
low values, since this would mean that the dark matter particle is
out of thermal equilibrium. Indeed, in a standard radiation
domination epoch, the thermalization rate is approximately
$\Gamma_{th}\sim \lambda^2 T$ when the temperature is way larger
than the Higgs mass, while when the temperature drops below the
Higgs mass, then $\Gamma_{th}\sim \lambda^2 T^5m_H^{-4}$. Thus the
ratio of the thermalization rate over the Hubble rate is maximized
when $T\sim m_W$, where $m_W$ is the mass of the W boson. The dark
matter particle thermalization condition down to the electroweak
epoch imposes the constraint $\lambda \geq
\sqrt{\frac{m_W}{M_p}}\sim 10^{-8}$ \cite{Burgess:2000yq}. The
caveat in the above considerations however \textit{is that the
axion is not a thermal relic} and is never at thermal equilibrium
with the particle content of the Universe, during all the stages
of the radiation domination era. Thermal relics are the WIMPs
only, which are in thermal equilibrium when the temperature is not
larger compared to their mass. When the temperature is larger than
the WIMPs masses, these decouple. When the WIMPs are produced,
they are produced by particles which are in thermal equilibrium
with their background, and the WIMPs energy spectrum is the same
with the particles that produces them in thermal equilibrium. On
the antipode of this physical picture, axions are not thermal
relics and are generated by coherent bosonic motion. The
difference between thermal and non-thermal relics are profound,
having different relic abundance, masses and of course couplings.
In the case of the axions, there is no restrictive lower bound to
the coupling, related with thermal equilibrium arguments. In fact,
a very weak sector is a desirable feature for the Higgs
phenomenology.

\begin{figure}
\centering
\includegraphics[width=20pc]{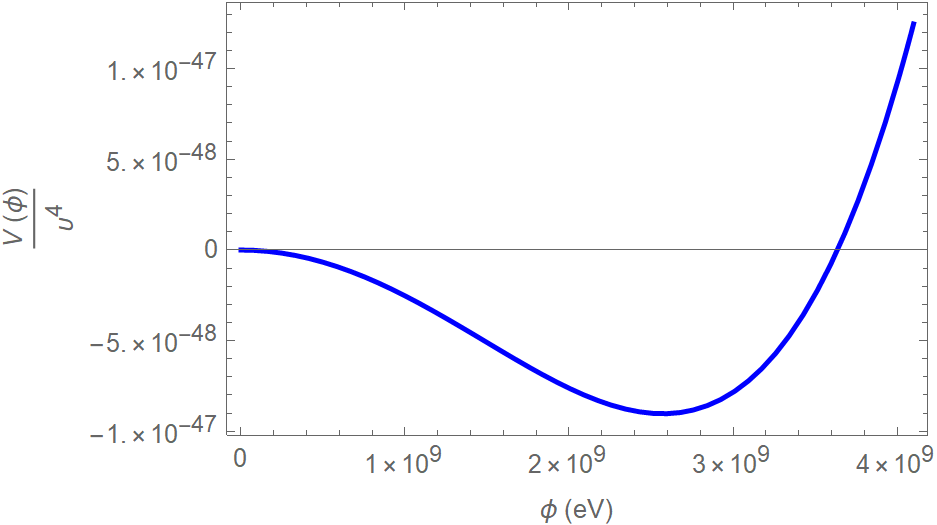}
\includegraphics[width=20pc]{higgsSB.png}
\caption{Tachyonic axion potential (left plot) and the Higgs
potential (right plot) after the electroweak symmetry
breaking.}\label{tachyonaxioncase}
\end{figure}

Let us proceed to the theory at hand, and the first thing to
discuss is the physics of the model at hand. During the
electroweak transition the Higgs particle acquires a vacuum
expectation value so $H=\frac{v+h+i h_1}{\sqrt{2}}$, therefore the
terms which couple the Higgs to the hidden axion sector modify the
tree order mass of the axion. Thus, although prior to the
electroweak breaking the axion was a normal particle, with
ordinary mass term, after the electroweak breaking the tree order
mass is of the form $m_{eff}^2= m_a^2-\lambda v^2$, thus depending
on the values of $\lambda$ it is highly possible that the axion
turns to a tachyon at the post-electroweak epoch. Indeed, this is
the scenario we shall consider. Obviously if $\lambda v^2\ll ma^2$
the axion never turns into a tachyon and the electroweak epoch
will just modify its tree order mass. We shall consider the case
in which $\lambda=\frac{2 m_a^2}{v^2}$, and since $v=246\,$GeV and
$m_a=10^{-10}$eV, it turns out that $\lambda\sim 10^{-43}$. For
this value, the tree order potential after the electroweak
breaking is well-bounded from below, since the inequality
$\lambda_H \frac{gv^2}{2M^2}\geq \lambda^2$ is satisfied. For the
choice of $\lambda$ we discussed above, the tree order potential
for the axion sector becomes,
\begin{equation}\label{higgsaxiontreepotential}
V(\phi)=-\frac{1}{2}m_a^2\phi^2+\frac{g v^2}{2 M^2}\phi^4 \, .
\end{equation}
Hence, the axion is basically a tachyon at tree order, and for the
values of the parameters chosen as indicated above, the tree order
potential can be found in Fig. \ref{tachyonaxioncase} and the
direct comparison with the Higgs potential can be found in the
same plot.
As it can be seen in Fig. \ref{tachyonaxioncase}, the tachyonic
axion potential has a minimum for a field space value different
from that at the origin. Thus the tachyonic axion rolls to the new
potential minimum. Once it reaches the new minimum, the vacuum
state $(h,\phi)=(0,v_s)$ decays to the much more deeper and
energetically favorable vacuum state $(h,\phi)=(v,0)$, in the way
described in the previous section. Thus the same physical picture
occurs in this case too, however in this case the axion is
practically a tachyon, so its a rather unappealing physical
situation. The effect on the gravitational waves though would be
the same, with the difference that a tachyonic instability drives
the short kination eras. Finally, the result does not change if
1-loop corrections are included in the tachyonic axion tree order
potential.

\section{An Idle Calculation for the Sake of Completeness: Inclusion of High Temperature Effects in the Axion Sector}

As we discussed earlier, the axion particle is not a thermal dark
matter relic, such as the WIMPs, but it is a non-thermal relic,
generated by coherent bosonic motion. Thus it is senseless to
consider high temperature effects on the axion potential at the
electroweak phase transition. However, for completeness and just
out of curiosity, in this section we shall consider high
temperature effects on the axion potential. As is probably
expected from the reader, the new minimum present at tree order
due to the higher dimensional operators in the axion potential
disappear at high temperature, only when 2-loop high-temperature
or all the higher order daisy graphs are included, and the new
minimum reappears once the temperature drops. Thus the physical
picture described in the previous sections, in which the axion
undergoes several swift kination eras is delayed if high
temperature effects are considered. But as we already mentioned,
this is an idle calculation, the high temperature effects never
affect the axion, because \textit{the axion is not a thermal
relic}.

\begin{figure}
\centering
\includegraphics[width=20pc]{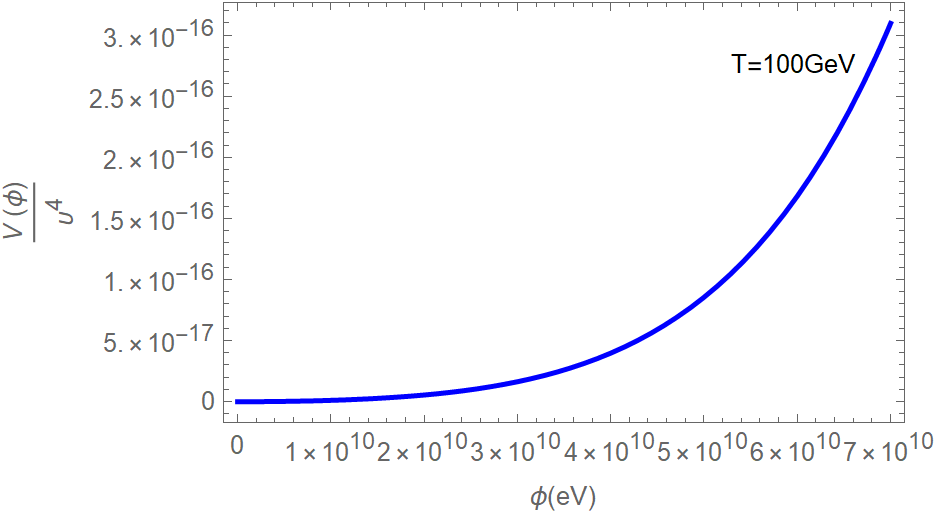}
\includegraphics[width=20pc]{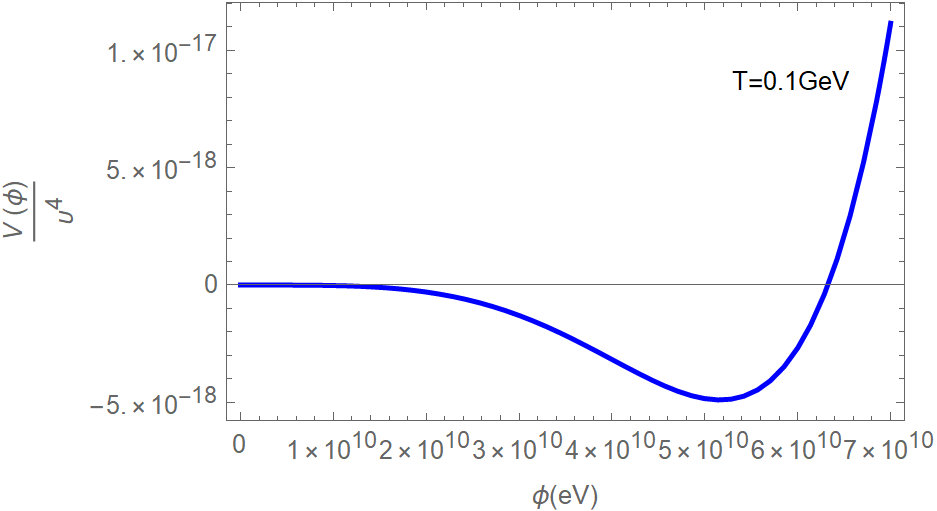}
\includegraphics[width=20pc]{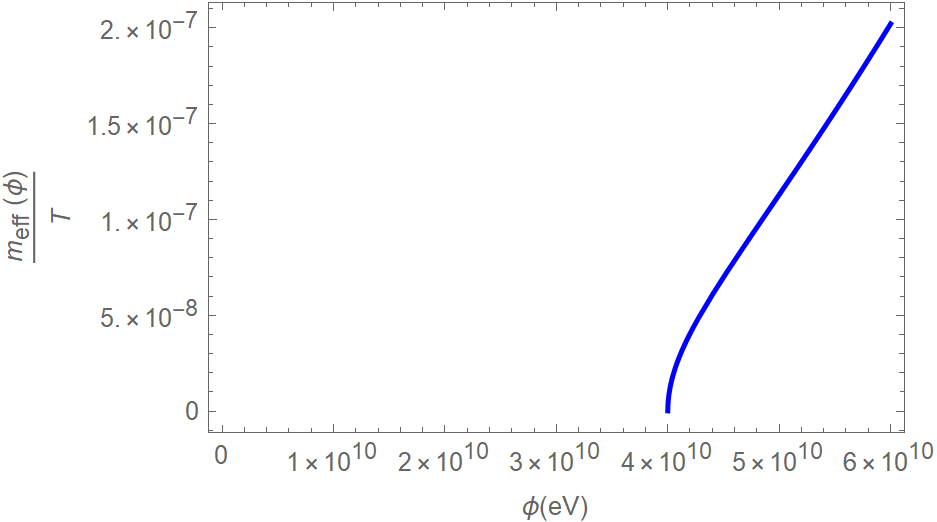}
\caption{The high temperature effective potential at 2-loops for
$T=100\,$GeV (upper left plot), the effective potential at
$T=0.1\,$GeV (upper right plot) and the behavior of
$\frac{m_{eff}(\phi)}{T}$  for all relevant field values (bottom
plot) for $T=100\,$GeV which also holds true for $T=0.1\,$GeV.
When the temperature is of the electroweak scale order $T\sim
100\,$GeV, the axion potential has the true minimum at the origin
(upper left plot). The axion potential develops a minimum only
when the temperature drops as low as $T\sim 0.1\,$GeV (upper right
plot).}\label{hightemp1loop68}
\end{figure}

\begin{figure}
\centering
\includegraphics[width=20pc]{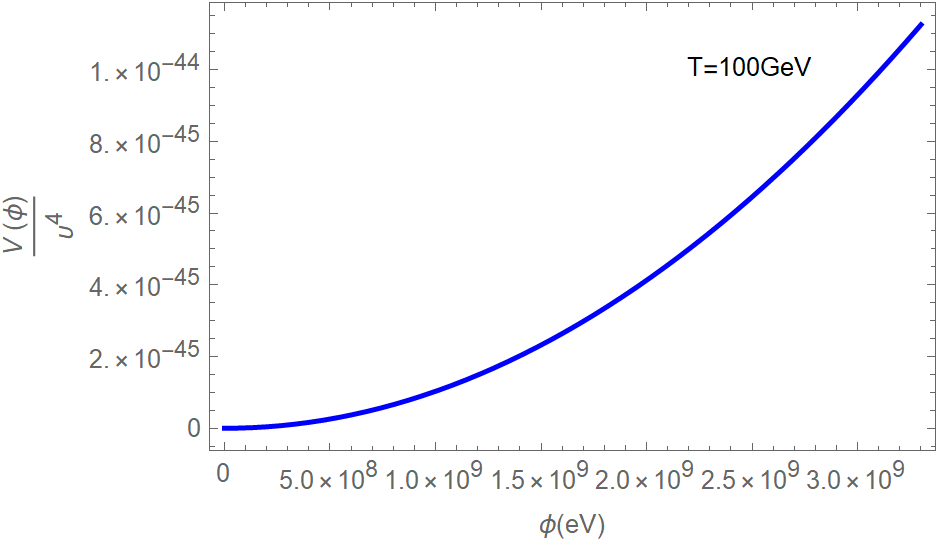}
\includegraphics[width=20pc]{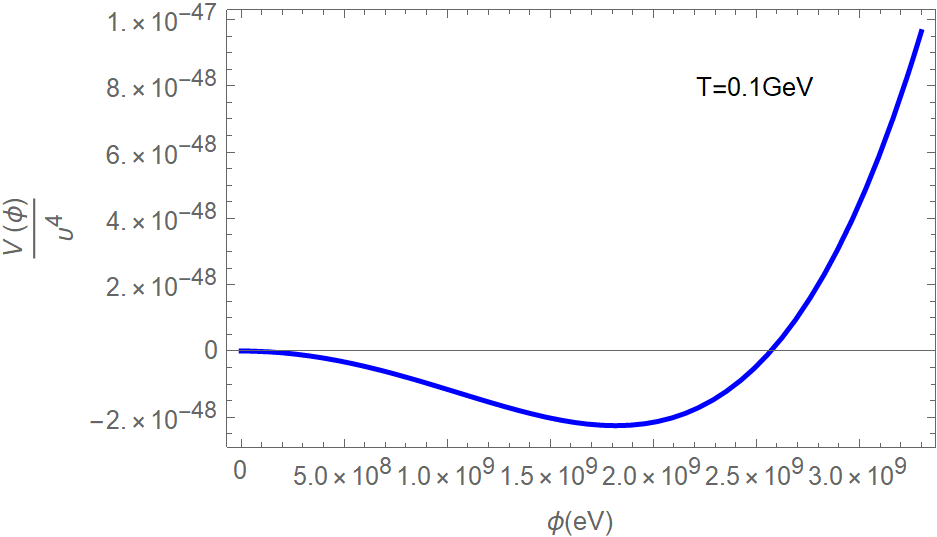}
\includegraphics[width=22pc]{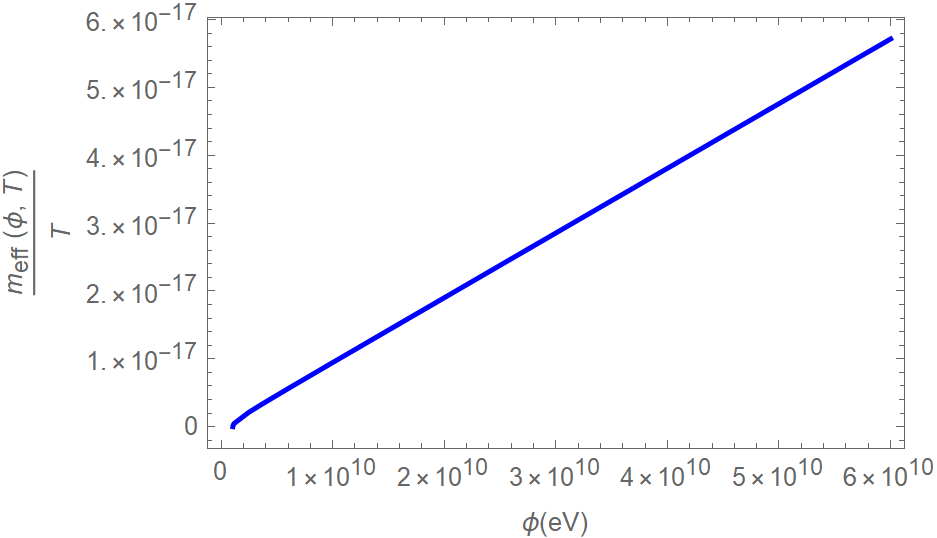}
\caption{The high temperature effective potential for all daisy
graphs, for the tachyonic axion case, for $T=100\,$GeV (upper left
plot), the daisy resumed effective potential at $T=0.1\,$GeV
(upper right plot) and the behavior of
$\frac{m_{eff}(\phi,T)}{T}$,  for all relevant field values
(bottom plot) for $T=100\,$GeV which also holds true for
$T=0.1\,$GeV. When the temperature is of the electroweak scale
order $T\sim 100\,$GeV, the tachyon axion potential has the true
minimum at the origin (upper left plot). The axion potential
develops a minimum only when the temperature drops as low as
$T\sim 0.1\,$GeV (upper right plot).}\label{tachyonaxiopot}
\end{figure}

Let us consider in brief the high temperature cases in this
section. We start with the physically appealing case in which the
new minimum in the axion potential is generated by the dimension
six and dimension eight operators. We shall consider one two loop
effects in this case. The high temperature corrected effective
potential of the axion at 1-loop, including zero-temperature
1-loop effects, has the following form \cite{Quiros:1999jp},
\begin{align}\label{newaxionpotential}
V_a^{1-loop T\neq
0}=\mathcal{V}_a(\phi)+V^{1-loop}(\phi)+\frac{T^4}{2\pi^2}\mathcal{J}_b\left(\frac{m_{eff}^2(\phi)}{T^2}
\right)\, ,
\end{align}
where the tree order potential $\mathcal{V}_a(\phi)$ and the zero
temperature 1-loop effective potential $V^{1-loop}(\phi)$ are
defined in Eqs. (\ref{axioneffective68}) and
(\ref{oneloopaxionzerotemperature}) respectively. Also the
effective mass of the axion $m_{eff}^2(\phi)=\frac{\partial^2
V(\phi,h)}{\partial \phi^2}$ is defined in Eq.
(\ref{axioneffectivemass}). Furthermore, the function
$\mathcal{J}_b\left(\frac{m_{eff}^2(\phi)}{T^2} \right)$ for small
value of $\frac{m_{eff}^2(\phi)}{T^2}$ \cite{Quiros:1999jp},
\begin{equation}\label{smallJ}
\mathcal{J}_b\left(x\right)\simeq
-\frac{\pi^4}{45}+\frac{\pi^2}{12}\frac{m_{eff}^2(\phi)}{T^2}-\frac{\pi}{6}\frac{m_{eff}^{3/2}(\phi)}{T^3}-\frac{1}{32}\frac{m_{eff}^4(\phi)}{T^4}\log\left(
\frac{m_{eff}^2(\phi)}{a_bT^2}\right)\, ,
\end{equation}
where $\log a_b=5.4076$. Thus adding all the contributions, we can
see that the effective mass-dependent logarithm terms cancel, and
the resulting effective potential in the limit
$\frac{m_{eff}^2(\phi)}{T^2}\ll 1$ reads at leading order,
\begin{align}\label{leadingorderhightemp68}
& V_a^{1-loop T\neq 0}\simeq m_a^2f_a^2\left(1-\cos
\left(\frac{\phi}{f_a}\right)
\right)-\lambda\frac{v^2\phi^4}{M^2}+g\frac{v^2\phi^6}{M^6}+\frac{m_{eff}^4(\phi)}{64\pi^2}\left(
\ln
\left(\frac{m_{eff}^2(\phi)}{\mu^2}\right)-\frac{3}{2}\right)\\
\notag &
+\frac{m_{eff}^2(\phi)}{24}T^2-\frac{T}{12\pi}\left(m_{eff}^2(\phi)
\right)^{3/2}+\frac{m_{eff}^4(\phi)}{64\pi^2}\left(\log(a_b)-\frac{3}{2}
\right)+\frac{m_{eff}^4(\phi)}{64\pi^2}\log \left(
\frac{T^2}{\mu^2}\right)\, ,
\end{align}
and as it proves, the last two terms do not affect the physics
corresponding to the above effective potential, when the
approximation $\frac{m_{eff}^2(\phi)}{T^2}\ll 1$ holds true. A
simple analysis for the above potential, by using the same values
for the free parameters as in the tree order potential, indicates
that when the temperature is of the order $T\sim 100\,$GeV, the
axion potential has the same minimum as the tree order potential
and this said behavior continues as the temperature drops. For all
the temperatures and field values, the approximation
$\frac{m_{eff}^2(\phi)}{T^2}\ll 1$ always holds true. This is
peculiar, so we included the 2-loop correction to the axion
effective potential at high temperature $\sim
\frac{gv^2}{2M^4}\phi^2 T^4$ \cite{Bodeker:2004ws}, and for
$\lambda\sim \mathcal{O}(10^{-10})$ an interesting behavior
occurs. Specifically, the axion minimum at $T\sim 100\,$GeV, is at
the origin, and when the temperature drops as low as $T\sim
0.1\,$GeV, the axion potential develops the second minimum which
is not (yet) identical to the zero-temperature one. The new
minimum becomes identical to the tree order minimum when the
temperature effects are no longer dominant. We need to note that
there is no barrier between the local maximum at the origin and
the new local minimum. Thus from the perspective of phase
transitions, this behavior mimics a second order phase transition.
For all the temperatures and field values, the approximation
$\frac{m_{eff}^2(\phi)}{T^2}\ll 1$ always holds true. This said
behavior is depicted in the three plots of Fig.
\ref{hightemp1loop68}, where we plot the high temperature
effective potential at 1-loop for $T=100\,$GeV (upper left plot),
the effective potential at $T=0.1\,$GeV (upper right plot) and the
behavior of $\frac{m_{eff}(\phi)}{T}$ for $T=100\,$GeV for all
relevant field values (bottom plot) which also holds true for
$T=0.1\,$GeV.
Regarding the tachyon axion case, we shall include all the higher
order loop graphs, the so-called daisy contributions at all
orders, and the tachyon axion effective potential in this case
reads,
\begin{equation}\label{leadingorderhightemptachyon}
V_a^{daisy T\neq 0}\simeq -\frac{1}{2}m_a^2\phi^2+\frac{g
v^2}{2M^2}\phi^4+\frac{g v^2}{2 M^2}T^2-\frac{1}{12
\pi}\left(-m_a^2+\frac{gv^2}{2M^2}T^2+\frac{6gv^2}{M^2}\phi^2
\right)^{3/2}T\, .
\end{equation}
In this case too, only when $g\sim\mathcal{O}(10^{-35})$, when the
temperature is as high as $T\sim 100\,$GeV, the minimum of the
tachyonic axion potential disappears, and reappears at $T\sim
0.1\,$GeV, as it can be seen in the left and right upper plots of
Fig. \ref{tachyonaxiopot}. In the bottom plot we present the
behavior of $m_eff(\phi,T)/T$, where $m_eff(\phi,T)=\frac{T^2
\left(g v^2\right)}{2 M^2}+\frac{12 \phi ^2 \left(g v^2\right)}{2
M^2}-m_a^2$.
As it can be seen the high temperature approximation holds true
for all the relevant field values, as long as the perturbation
theory approximation is not violated.

Thus, in the case the high temperature corrections are included,
the qualitative picture of the tree order case does not change, it
is only delayed.

\section{Conclusions and Discussion}

In this work we considered the effects of short axion kination
eras on the energy spectrum of the primordial gravitational waves
for modes re-entering the Hubble horizon near the electroweak
symmetry breaking epoch and beyond, well inside the radiation
domination era. We assumed that the axion particle is coupled to
the Higgs particle solely via higher dimensional operators, of
dimension six and eight. When the electroweak symmetry breaking
epoch occurred, these higher dimensional operators modified the
axion potential, causing a new minimum to it and thus
destabilizing the minimum at the origin. Once the axion
oscillations at the origin are destabilized after some time, the
axion starts to swiftly roll its potential heading to the new
minimum, experiencing a kinetic evolution, with its energy density
redshifting as $\rho_a\sim a^{-6}$. Depending on whether the axion
composes the whole or some part of the dark matter, this kinetic
evolution may affect to some extent the total background EoS,
changing it from the radiation value $w=1/3$ to some value closer
to the kination domination value $w=1$. We chose a conservative
approach and assumed that the background EoS parameter value
changed to $w=0.5$. When the axion reached its new potential
minimum, due to the fact that the Higgs electroweak vacuum is
energetically more favorable than the axion minimum, the latter
decays to the Higgs vacuum and the Universe is again described by
the latter. Thus the axion returns to the origin of its potential,
and due to the quantum fluctuations it starts its oscillations,
due to the dominance of the potential term $\sim \phi^2$. After
some considerable time the oscillations are disrupted and the
axion starts to roll towards in its new minimum and the procedure
we described is repeated again. Thus during the radiation
domination era these short axion kination epochs may occur many
times. These deformations of the background EoS parameter may have
measurable effects on the primordial gravitational waves. These
effects mainly depend on how many times the short in duration
axion kination epochs occur, on the theory that describes the
inflationary era, the background EoS parameter value during the
short kination era and the reheating temperature. We investigated
several scenarios, disregarding the primordial gravitational waves
coming from the electroweak phase transition, and the results we
found indicate a characteristic structure in the energy spectrum
which can be observed in future gravitational wave experiments.
Finally, we considered an alternative scenario with a direct
renormalizable coupling between the Higgs and the axion. In this
scenario, the axion is rendered a tachyon during and beyond the
electroweak breaking epoch, but the same physical behavior occurs
as in the case of the dimensions six and dimension eight
non-renormalizable operators. Finally, we included a purely
academic study, including high temperature effects in the axion
potential. This is not physically justified though, since the
axion never thermalizes with its background, since it is a
non-thermal relic which is generated by coherent bosonic
oscillations. However, just out of curiosity we included this
study, and as we demonstrated, the temperature effects delay the
commencing of the short kination eras, but the same physics occur
nonetheless. We also discussed the effects of this axion rolling
scenario on the matter domination epoch. During the latter, the
effects would be more dramatic, and we examined the case that the
axion does not roll swiftly from the beginning of its motion, but
it slow-rolls initially. This would cause multiple short early
dark energy epochs during the matter domination epoch. A future
perspective could be to include several other singlet axion-like
particles in this scenario and investigate the overall effect on
the short kination eras, due to the electroweak symmetry breaking.
Apparently, the whole scenario is based on the fact that the
reheating temperature was higher than $T\sim 100\,$GeV, and that
the electroweak epoch occurred during the radiation domination
era. However, this is not certain, since the reheating era may
have not been so large \cite{Hasegawa:2019jsa}. Thus one should
consider alternative scenarios that may control the electroweak
symmetry breaking epoch. We hope we shall address these aspects in
some future works. Finally, before closing there are a lot to say
about the shape of the predicted gravitational wave energy
spectrums of different models, regarding the slope and the peak
frequency. There are a lot to say at this point, on how to
discriminate these curves and pin point the correct model, and
also one must also take into account the direct effect of the
reheating temperature, the frequency range that the peak is
observed, how wide the peak is, is the peak and curve observed by
one or multiple detectors, and how would the effects of phenomena
we ignored, like primordial gravitational waves or the effects of
a first order phase transition would modify the above picture.
Undoubtedly, these will be very important issues that
theoreticians should discuss and collaborate on, before the LISA
and Einstein telescope yield their first observational data on
primordial stochastic gravitational waves.

\section*{Acknowledgments}

This research has been is funded by the Committee of Science of
the Ministry of Education and Science of the Republic of
Kazakhstan (Grant No. AP14869238)

\end{document}